\documentclass[twocolumn,amsmath,amssymb,prb,longbibliography]{revtex4-2}

\usepackage{graphicx}
\usepackage{dcolumn}
\usepackage{bm}
\usepackage{color}
\usepackage{soul}
\usepackage{notes2bib}
\usepackage[binary-units=true]{siunitx}

\begin{document}



\title{Ultralong-term high-density data storage with atomic defects in SiC}

\author{M.~Hollenbach$^{1}$}
\author{C.~Kasper$^{2}$}
\author{D.~Erb$^{1}$} 
\author{L.~Bischoff$^{1}$} 
\author{G.~Hlawacek$^{1}$} 
\author{H.~Kraus$^{3}$}
\author{W.~Kada$^{4}$} 
\author{T.~Ohshima$^{5,7}$} 
\author{M.~Helm$^{1,6}$}
\author{S.~Facsko$^{1}$}
\author{V.~Dyakonov$^{2}$}
\author{G.~V.~Astakhov$^{1}$}
\email[E-mail:~]{g.astakhov@hzdr.de}

\affiliation{$^{1}$Institute of Ion Beam Physics and Materials Research, Helmholtz-Zentrum Dresden-Rossendorf, 01328 Dresden, Germany \\
$^2$Experimental Physics 6 and W\"{u}rzburg-Dresden Cluster of Excellence ct.qmat, Julius-Maximilian University of W\"{u}rzburg, 97074 W\"{u}rzburg, Germany \\ 
$^3$Jet Propulsion Laboratory, California Institute of Technology, Pasadena, CA 91109, USA \\
$^4$Faculty of Science and Technology, Gunma University, Kiryu, Gunma 376-8515, Japan \\
$^5$National Institutes for Quantum Science and Technology, Takasaki, Gunma 370-1292, Japan \\ 
$^6$Technische Universit\"{a}t Dresden, 01062 Dresden, Germany \\ 
$^7$Department of Materials Science, Tohoku University, 6-6-02 Aramaki-Aza, Aoba-ku, Sendai 980-8579, Japan}

\begin{abstract}
There is an urgent need to increase the global data storage capacity, as current approaches lag behind the exponential growth of data generation driven by the Internet, social media and cloud technologies. In addition to increasing storage density, new solutions should provide long-term data archiving that goes far beyond traditional magnetic memory, optical disks and solid-state drives. Here, we propose a concept of energy-efficient, ultralong, high-density data archiving based on optically active atomic-size defects in a radiation resistance material, silicon carbide (SiC). The information is written in these defects by focused ion beams and read using photoluminescence or cathodoluminescence. The temperature-dependent deactivation of these defects suggests a retention time minimum over a few generations under ambient conditions. With near-infrared laser excitation,  grayscale encoding and multi-layer data storage, the areal density corresponds to that of Blu-ray discs. Furthermore, we demonstrate that the areal density limitation of conventional optical data storage media due to the light diffraction can be overcome by focused electron-beam excitation.  
\end{abstract}

\date{\today}

\maketitle

\section{Introduction} 

In 2012, the amount of digital data in the world surpassed one Zettabyte (ZB or $10^{21}$ bytes), indicating the dawn of the Zettabyte Era \cite{10.1007/s11390-014-1420-2}. The globally-generated data are continuously  increasing and expected to be above 100~ZB/year by 2025 \cite{10.1038/s41467-021-22687-y}. The energy consumption of data centers amounts about 1\% of global electricity and this number can increase to 3--13\% by 2030 \cite{10.3390/challe6010117}. The limited storage time requires data migration within several years to avoid any data loss. This substantially increases the energy consumption, because a significant amount of energy is consumed during such a data migration \cite{10.1038/lsa.2014.58s5d}. The demand for reducing energy consumption, prolonging storage time and increasing capacities for big data has led to a continuous improvement of current technologies such as magnetic media, optical disks and solid-state drives. 

Magnetic memory is the primary choice for data archiving because of its large storage capacity. A recently developed prototype magnetic tape shows a recording performance with an areal storage density of $317.3 \, \mathrm{Gbit / in^2}$ \cite{10.1109/tmag.2021.3076868}. 
The increase of storage density requires the decrease of the magnetic particle size. In this case, thermal fluctuations and diffusion processes become significant, resulting in a decrease of storage time. This can  be overcome using  materials with higher coercivity, which in turn lead to a higher energy required to store one magnetic bit. On the other hand, optical discs can provide seemingly unlimited storage time \cite{10.1103/physrevlett.112.033901}. However, the diffraction limit restricts the smallest recording bit to half of the light wavelength, which in turn limits the maximum storage capacity. Using multidimensional optical recording (i.e., multiple layers inside the medium, spectral domain, grayscale encoding etc.), the storage capacity per volume can be drastically increased \cite{10.1038/nature08053, 10.1002/adma.201002413, 10.1038/srep26163}. In the demonstrated concepts, the multi-dimensional optical recording is based on the material modification using ultrashort laser pulses \cite{10.1117/12.2220600, 10.1038/s41467-018-03589-y, 10.1364/optica.433765, 10.1002/adma.201002413}. The energy required to write a single bit in these approaches is above $\mathrm{100 \, nJ / bit}$. This value is many orders of magnitude higher than for  magnetic disk drives and solid-state drives, which is in the range of  $\mathrm{1 \, pJ / bit}$ and $\mathrm{100 \, fJ / bit}$, respectively \cite{10.1109/mascot.2008.4770567, 10.1088/0022-3727/46/7/074003}. These values can be further improved by reducing the heat load \cite{10.1038/nature20807, 10.1088/0022-3727/46/7/074003}. On the other hand, the optical data storage does not need to be refreshed and hence does not consume any energy in the idle state. In case of long-term data archiving, there is a trade-off between the energy needed to write a single bit and the storage lifetime \cite{10.1038/lsa.2014.58s5d}. 

Other approaches for ultrahigh-density data storage are under development.  Particularly, DNA digital data storage provides the highest storage density per volume \cite{10.1126/science.1226355}. 
 In an alternative approach, atomic-size memories have the highest areal storage density \cite{10.1038/s41467-018-05171-y}. However,  times for writing and reading in these two cases are currently impractical. Color centers in the solid state constitute a particularly robust type of atomic-scale storage. Indeed, data storage using the nitrogen-vacancy (NV) centers in diamond has been demonstrated \cite{10.1126/sciadv.1600911}. The bit encoding is realized using charge-state control with laser light and the readout is performed via photoluminescence  (PL). However, the bit patterns are stored within the NV centers for only over a week in the dark, which is not sufficient for practical use. 

Here, we propose a concept of long-term WORM (write once read many) data storage based on SiC, as schematically shown in Fig.~\ref{fig1}a.  SiC hosts atomic-scale color centers, especially the silicon vacancy ($\mathrm{V_{Si}}$), i.e., the absence of silicon atom in the lattice site. It possesses a room-temperature broad PL band in the near infrared (NIR) spectral range. The $\mathrm{V_{Si}}$ defects are created by a focused proton or helium ion beam, providing high spatial resolution, fast writing speed and low energy for storing a single bit. The diffraction limit of storage density inherent to optical media is overcome by 4D encoding schemes. In these schemes, the three spatial dimensions and additional fourth intensity dimension (grayscale encoding) are realized by controlling the lateral position and depth as well as the number of $\mathrm{V_{Si}}$ defects through the ion energy and fluence, respectively. A scanning PL confocal microscope is used to optically read out the stored data. Alternatively, the use of cathodoluminescence (CL) instead of PL  can dramatically improve the areal storage density.

 \section{Results and Discussion} 

\begin{figure}[t]
\includegraphics[width=.45\textwidth]{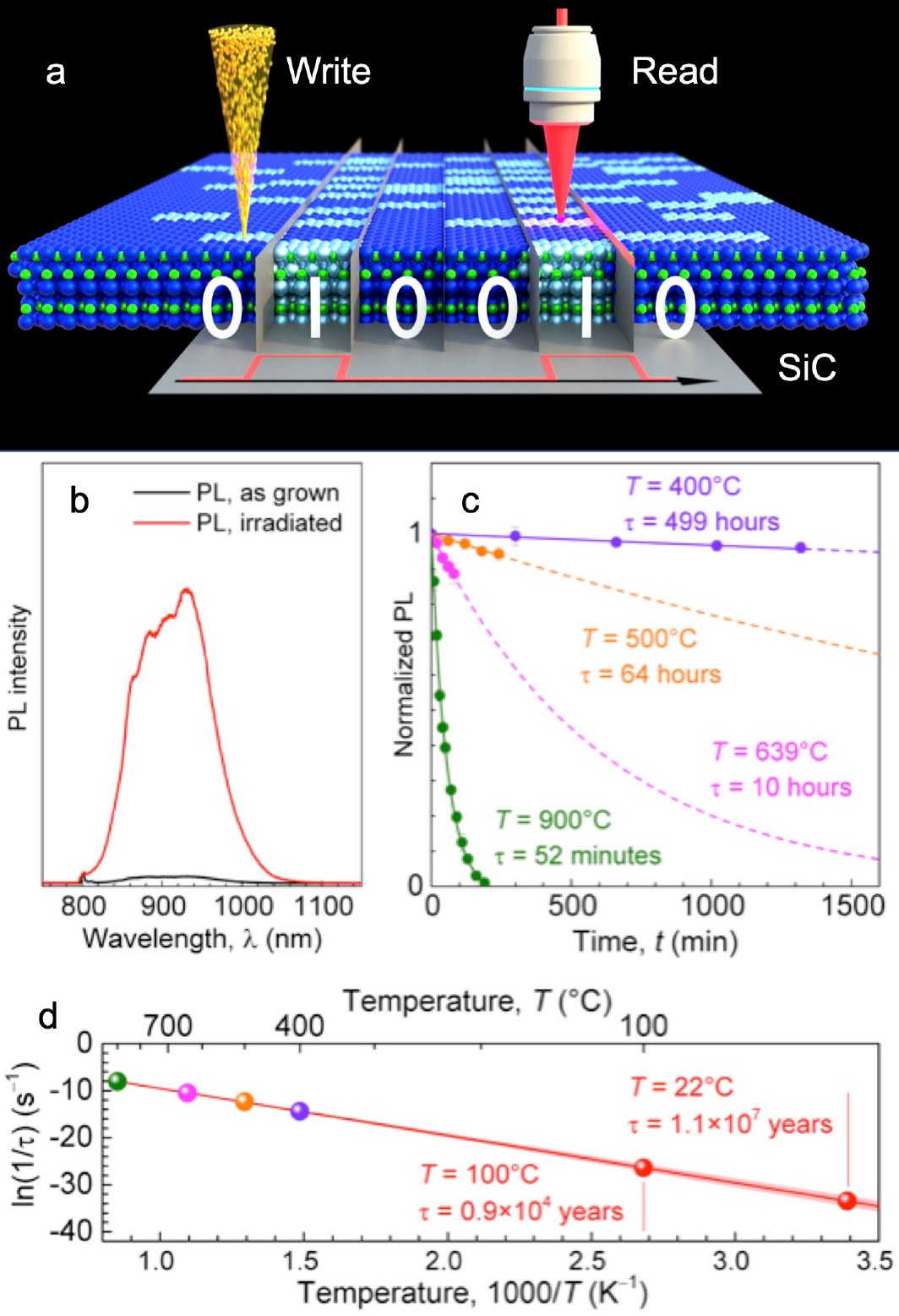}
\caption{A concept of long-term optical data storage in SiC. a) The information is written in optically active atomic defects by a focused ion beam and read using the defect PL or CL. The excitation wavelength is  $785 \, \mathrm{nm}$.  b) Typical PL spectrum in as-grown HPSI SiC and after ion irradiation.  c)  The decay of the $\mathrm{V_{Si}}$ PL intensity at different 
temperatures. The solid lines are fits to $\exp (-t / \tau)$ and the dashed lines represent extrapolation to longer annealing time $t$. d)  Arrhenius plot of the PL decay rate $1/ \tau$.  The shaded area around the solid line represents the tolerance of the extrapolated values. At room temperature, the information in $\mathrm{V_{Si}}$ can be stored for more than ten million years. } \label{fig1}
\end{figure}

For our studies, we use commercial 4H-SiC wafers. To create atomic vacancies, we perform proton irradiation with a fluence of $1 \times 10^{15}\, \mathrm{cm^{-2}}$. We then measure  PL from  $\mathrm{V_{Si}}$ in the spectral range from $800$ to $1000 \, \mathrm{nm}$ under optimum excitation at $785 \, \mathrm{nm}$ \cite{10.1063/1.4870456}. Room-temperature PL spectra in as-grown sample and measured after irradiation are shown in Fig.~\ref{fig1}b, indicating clear activation of optically active $\mathrm{V_{Si}}$ defects without any post-irradiation treatment. In addition, we perform optically detected magnetic resonance (ODMR) experiments from  implanted areas (Supporting Information). The observation of the ODMR peak at $70 \, \mathrm{MHz}$ is a clear indication of the $\mathrm{V_{Si}}$ defects in 4H-SiC  \cite{PhysRevX.6.031014}. 

 \subsection{Storage time} 

It is well established that the optimum annealing temperature to heal the lattice from irradiation damage and simultaneously preserve $\mathrm{V_{Si}}$ is in the range $500 - 600^{\circ}\mathrm{C}$ \cite{10.1038/ncomms8578}. At higher annealing temperatures, the $\mathrm{V_{Si}}$ defects are rapidly removed, however, systematic studies have not been performed so far. Therefore, we  investigate  in detail the dependence of the integrated PL intensity on the annealing duration time $t$ for different annealing temperatures $T$ (Fig.~\ref{fig1}c).

To heal the crystal from the irradiation damage, we start from low annealing temperatures in steps of $100^{\circ}\mathrm{C}$ and duration of $90 \, \mathrm{min}$ at each step. It leads to initial  increase of the PL intensity by approximately 52\%. When we observe the PL starts to drop down at $640^{\circ}\mathrm{C}$, we stop to increase the temperature and perform several series of annealing experiments at lower temperatures by varying the annealing time. At the beginning of each temperature series, suitable temperature values and associated annealing times are determined by the test run. At the final step, we perform annealing at a higher temperature of $900^{\circ}\mathrm{C}$ until the complete PL disappearance and confirm mono-exponential decay. To increase the precision and ensure the repeatability of the relative PL change, all annealing measurement are performed on the the same sample and from the same area. The complete annealing history of the sample is presented in Supporting Information. 

The general tendency is that the lower annealing temperature the  longer annealing time is necessary to observe any PL change (Fig.~\ref{fig1}c). Remarkably, even at  $T = 400^{\circ}\mathrm{C}$, a very small but nonzero decrease in the PL intensity is detectable  for long annealing time of 6 hours. The highest temperature $T = 900^{\circ}\mathrm{C}$ is chosen as the last of four annealing series, so that short annealing times in the range of minutes are needed to observe significant PL decay. Finally, the experiments in this series are carried out until  no  optically active $\mathrm{V_{Si}}$ defects are observed anymore. The annealing at $900^{\circ}\mathrm{C}$ shows a complete exponential decay, which can be fitted to $\exp (-t / \tau)$, as shown by the solid line in Fig.~\ref{fig1}b. For lower annealing temperatures, the PL change is small and the fit is given by the first term of the exponent Taylor expansion  $1-t / \tau$. 

The temperature dependence of the PL decay time $\tau$ is determined by the activation energy $E_a$, in accord with the Arrhenius law  \cite{10.1103/physrevlett.112.033901}
\begin{equation}
\frac{1}{\tau}  = A \exp \left(  - \frac{E_a}{k_B T} \right)   \,.
\label{Arrhenius}
\end{equation}
Here, $k_B$ is the Boltzmann constant. The  Arrhenius plot of $1 / \tau$  as a function of temperature is presented in Fig.~\ref{fig1}d. Form the fit to Eq.~(\ref{Arrhenius}), we obtain $E_a = 860 \pm 10 \, \mathrm{meV}$. For comparison, the thermal energy at room temperature is $26 \, \mathrm{meV}$. With the frequency factor $A = 1.6 \pm 0.1 \, \mathrm{Hz}$, we extrapolate the PL decay time to $\tau  =  10^7 \, \mathrm{years}$ at room temperature $T =22^{\circ}\mathrm{C}$ and $\tau  =  10^4 \, \mathrm{years}$ at the water boiling point under standard  pressure $T =100^{\circ}\mathrm{C}$. The extrapolation over such a  long time results in a very large error. But obviously, the information medium based on the $\mathrm{V_{Si}}$ defects in SiC can preserve  the information over a few generations, i.e. a minimum of a few hundred years.

 \subsection{Data storage with grayscale encoding} 

\begin{figure}[t]
\includegraphics[width=.47\textwidth]{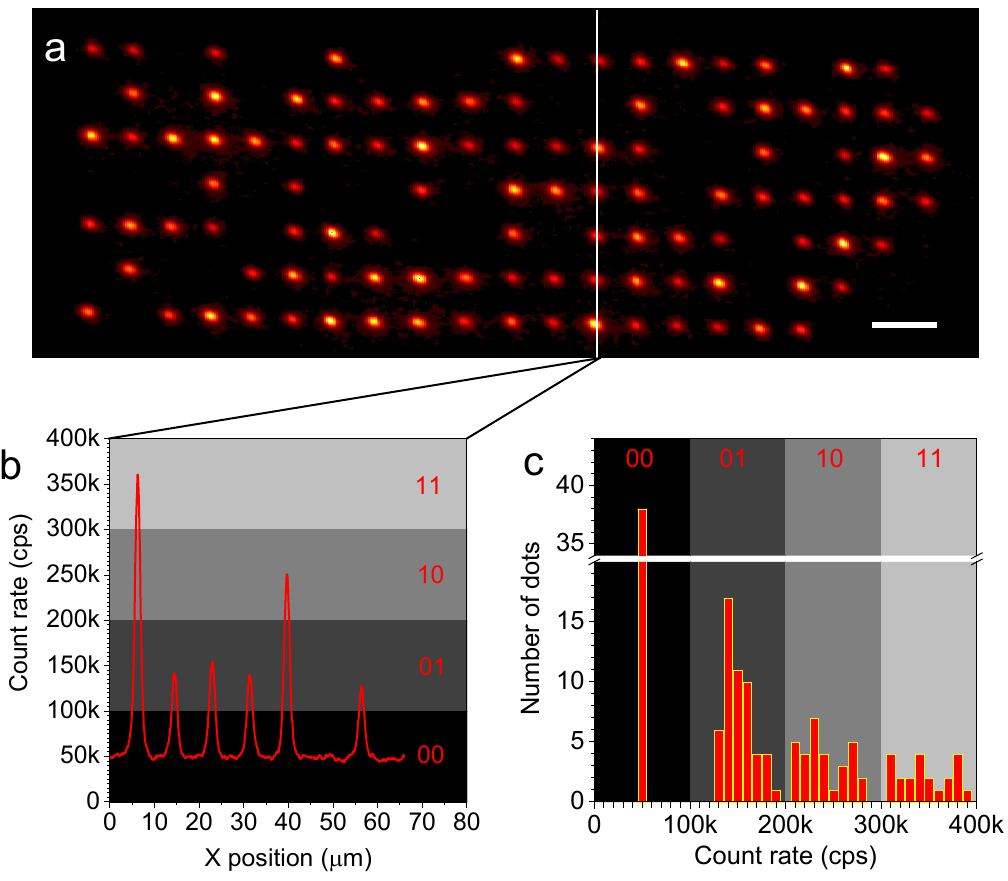}
\caption{Optical data storage in SiC using grayscale encoding. a) An example of the digital stream with two-bit encoding per pixel. The scale bar is $10 \, \mathrm{\mu m}$. b) Photon count rate of the $\mathrm{V_{Si}}$ PL recorded along one of the data tracks.  Four count rate levels represent the two-bit logical states 00, 01, 10 and 11. c)  Statistics of the PL count rate over all pixels.} \label{fig2}
\end{figure}

Next, we record the text "E pur si muove - Albeit it does move"  from Galileo Galilei \cite{baretti1757italian} in optically active $\mathrm{V_{Si}}$ defects. To perform digital recording, we convert this text to the hexadecimal representation of ASCII code, as described  in Supporting Information.  We create $\mathrm{V_{Si}}$ defects  using a proton microbeam focused to a round spot of approximately $1 \, \mathrm{\mu m}$ in diameter \cite{10.1021/acs.nanolett.6b05395}. A two-bit encoding scheme is used for each implantation pixel. We implant $4 \times 10^6$ protons  to write 01 and, correspondingly, double  and triple of this value to  write  10 and 11. The absence of implantation corresponds  to 00. 

To retrieve the recorded digital stream, we use  a home-built scanning confocal microscope as presented in  Fig.~\ref{fig2}a. The $\mathrm{V_{Si}}$ defects are excited with a 785-nm laser and the integrated PL is measured with a silicon single photon detector after filtering with a 800-nm longpass filter. Figure~\ref{fig2}b shows the $\mathrm{V_{Si}}$  PL along one of the data tracks, where the different count rates for the four logical states 00, 01, 10 and 11 are clearly seen. Figure~\ref{fig2}c shows the PL count rate statistics for all pixels. Four maxima can be seen which we identify with the four logical states. Finally we restore the text  "E pur si \textbf{i}uove - Albe\textbf{Y}t it does move".  Out of 294 recorded bits, we detect only 2 bit errors, leading  to two wrong characters (highlighted  in  bold). These errors are due to some variation of the PL intensity and can be corrected using redundant bits as a reference for count rates. 

\begin{figure}[t]
\includegraphics[width=.47\textwidth]{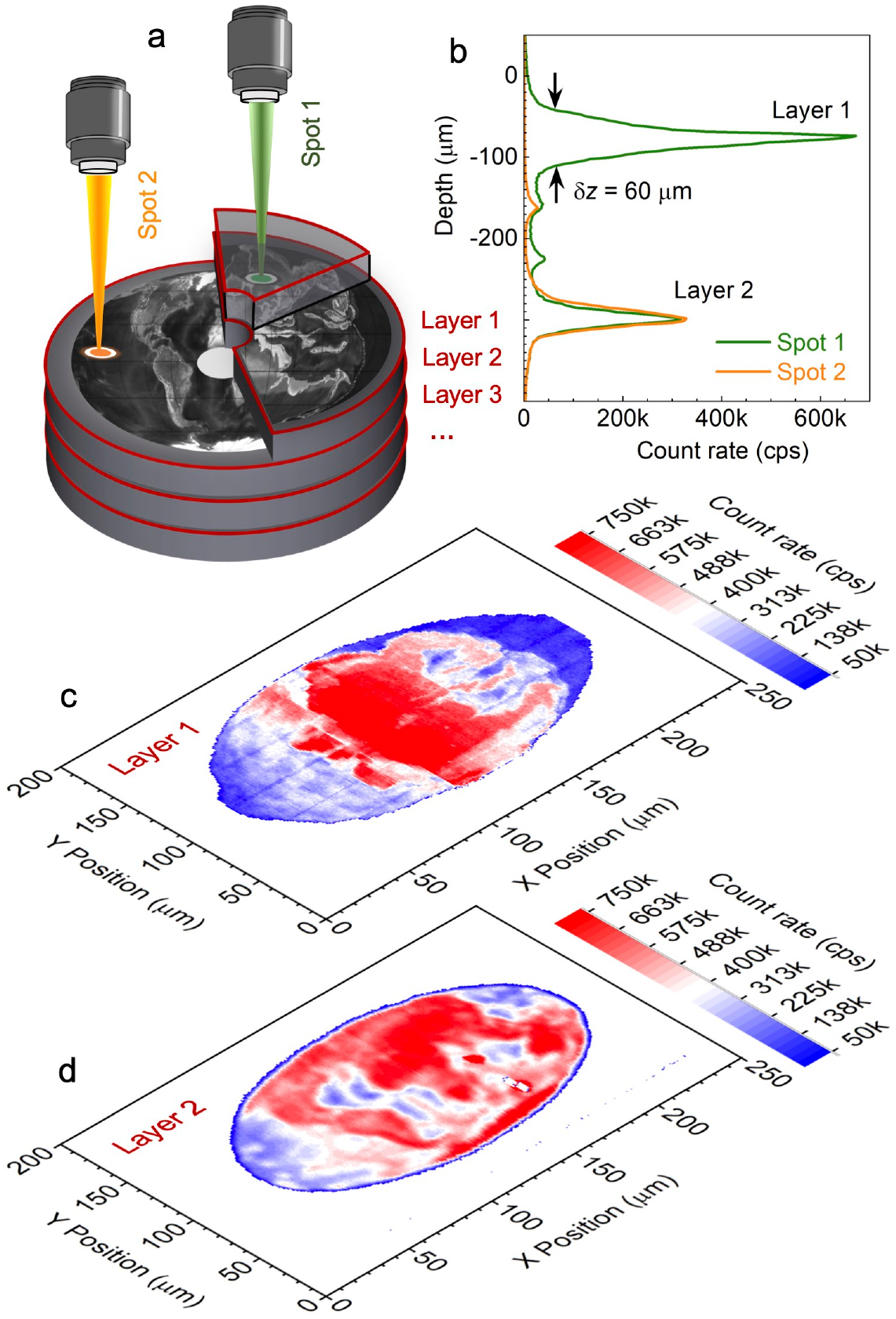}
\caption{Optical data storage in SiC using multi-layer encoding. a) A scheme of the optical storage medium with data layers of about $250 \, \mathrm{nm}$ thickness, separated by transparent spacers.  b) In-depth confocal PL scans through two spots. The spots are selected such that the irradiation fluence is the same in layer~2 and different in layer~1.  Due to the negligible  absorption of the excitation laser and $\mathrm{V_{Si}}$ PL in SiC, there is no cross-talking between the layers. c) A map of the Earth 240 million years ago stored in / retrieved from layer~1. An 8-bit grayscale encoding is used by the ion-beam writing and a 3-bit encoding is reinstated by the optical reading.  d) A map of the Earth nowadays stored in / retrieved from  layer~2. The background count rate of $50\mathrm{k} \, \mathrm{cps}$  caused by residual  $\mathrm{V_{Si}}$ is subtracted in c) and d). The  count rate of $88\mathrm{k} \, \mathrm{cps}$ corresponds to one level of the 3-bit grayscale encoding.} \label{fig3}
\end{figure}

To increase the storage capacity, we demonstrate multi-layer encoding as schematically  presented in Fig.~\ref{fig3}a. We use commercial SiC wafers  and record maps of the Earth  at different geological times. First, these maps are converted to 8-bit grayscale images (Supporting Information) with a distance between pixels of $80 \, \mathrm{nm}$. He ions with an energy of $25 \, \mathrm{keV}$ are focused to  a spot size of approximately $1 \, \mathrm{nm}$ on the sample surface in a helium ion microscope (HIM).   
To find the optimal recording conditions, we perform a series of experiments with different implantation fluences \cite{10.1103/PhysRevApplied.13.044054}, as described in Supporting Information. We find that the $\mathrm{V_{Si}}$ PL intensity increases nearly linearly up to a fluence of $\Phi_{max} = 1   \times 10 ^{13} \, \mathrm{cm^{-2}}$. It saturates at $5   \times 10 ^{13} \, \mathrm{cm^{-2}}$ and then drops down due to ion-induced damage of the SiC crystalline structure \cite{10.1007/s41871-020-00061-8}. The grayscale for focused ion-beam writing \cite{10.1038/s41467-022-35051-5}  is chosen such that  the maximum logical 8-bit state  is coded with $ \Phi_{max} $ and other bit states scale down linearly with the implantation fluence. 

After writing individual Earth maps in separated SiC wafers, they are stacked together (Fig.~\ref{fig3}a). We then use a confocal setup to retrieve the stored maps. In-depth scans through different points are shown in Fig.~\ref{fig3}b. The transparent spacer in these scans is given by the thickness of the wafer of  $250 \, \mathrm{\mu m}$. Considering the depth resolution of $60 \, \mathrm{\mu m}$ together with  the  working distance of the objective of $700 \, \mathrm{\mu m}$, we conclude that up to 10  layers can be retrieved without cross-talking. This is because of the weak absorption coefficient of SiC at the excitation and $\mathrm{V_{Si}}$ PL wavelengths \cite{10.1103/PhysRevApplied.9.054006}, which are longer than $380 \, \mathrm{nm}$ corresponding to the 4H-SiC band gap. The maps retrieved from two layers separated by  a spacer are presented in Fig.~\ref{fig3}c and d. Here, we use 3-bit encoding to reconstruct the maps.  

We now estimate the storage density limit of our 4D recording.  The lateral optical  resolution is given by the Rayleigh criterion $1.22 \lambda  /  \mathrm{NA}$, where $\lambda = 785 \, \mathrm{nm}$ is the excitation wavelength and $\mathrm{NA} = 0.81$ is the numerical aperture of the objective. We then consider 10 layers and 3-bit grayscale encoding for the third and  fourth dimension,  respectively. Finally, we obtain a storage density of $75 \, \mathrm{Gbit / in^2}$ for 10 layers.

\subsection{Cathodoluminescence readout} 

\begin{figure}[t]
\includegraphics[width=.48\textwidth]{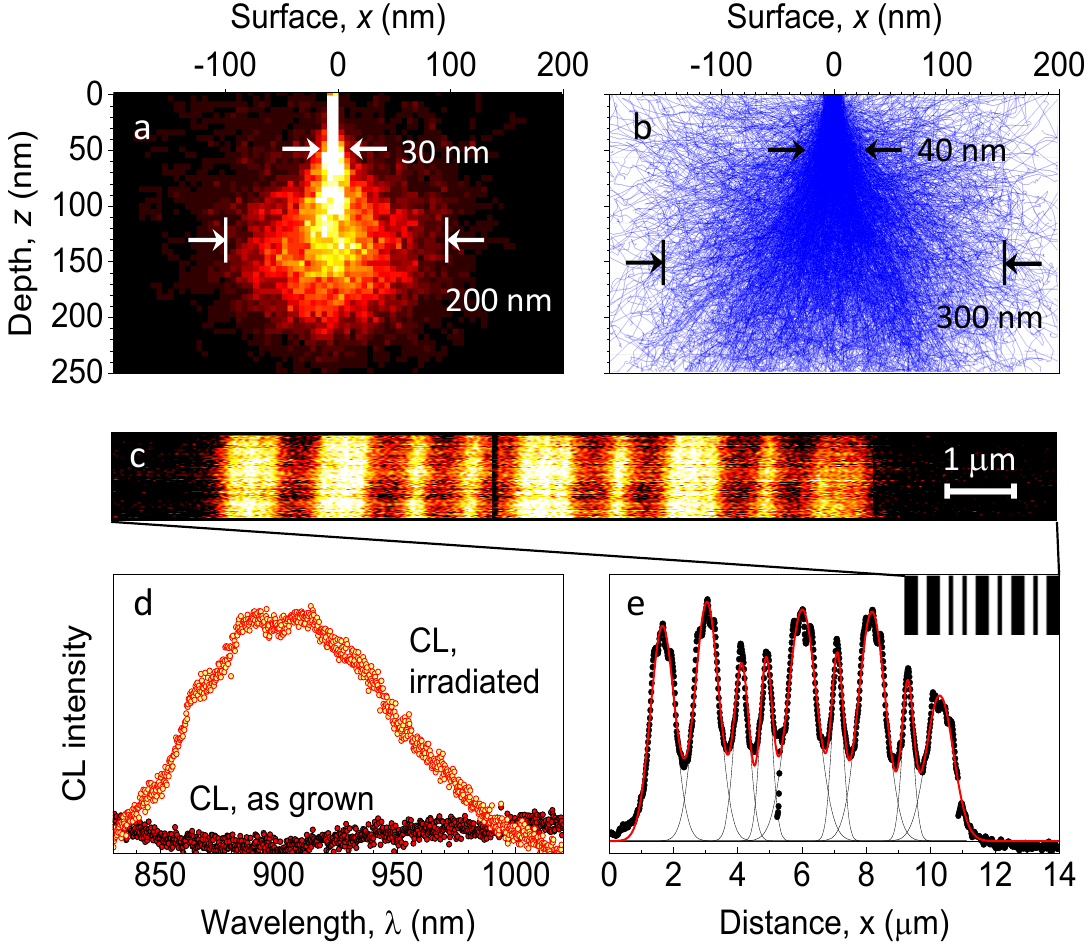}
\caption{Data storage in SiC with CL readout. a) SRIM simulation of the $\mathrm{V_{Si}}$ creation with He implantation with a landing energy of $25 \, \mathrm{keV}$ indicating longitudinal and lateral straggling. b) CASINO simulation of the longitudinal and lateral straggling of electrons with an energy of $5 \, \mathrm{keV}$. c) CL intensity map of the barcode written with HIM. d) CL spectrum from the irradiated and as-grown areas. e)  Integrated CL line scan through the barcode stored in c). The solid lines are a multi-gaussian fit. Insert shows the barcode stored in c). } \label{fig4}
\end{figure}

The diffraction limit of the minimum spot size under optical excitation is much larger than the implanted area in HIM. Figure~\ref{fig4}a shows a SRIM (Stopping and Range of Ions in Matter) simulation of the lateral and in-depth distribution of the $\mathrm{V_{Si}}$ defects created by He ions with  an energy of $25 \, \mathrm{keV}$. The penetration depth of He ions with this energy is about $250 \, \mathrm{nm}$. 
Due to lateral ion straggling inside the SiC layer, the implanted area is about $\sigma_{\mathrm{HIM}} \approx 200 \, \mathrm{nm}$ in diameter.  To increase the spatial resolution, we propose to use CL instead of PL. Figure~\ref{fig4}b shows straggling of electrons with an energy of $5 \, \mathrm{keV}$ in SiC, simulated with CASINO software (monte CArlo SImulation of electroN trajectory in sOlids). The lateral straggling is about $\sigma_{\mathrm{CL}} \approx 300 \, \mathrm{nm}$. Though electrons with this energy penetrate in SiC over several $\mathrm{\mu m}$, the simulation of the depth is limited to a depth of $250 \, \mathrm{nm}$, which is the thickness of the epitaxial SiC layer used in our CL experiments.   

CL spectra in the range from $800$ to $1000 \, \mathrm{nm}$ are shown in  Fig.~\ref{fig4}d.  There is a vanishingly low CL signal in  an as-grown  sample compared  to PL. The reason for that is a vanishingly low concentration of $\mathrm{V_{Si}}$ defects in epitaxial layers compared to HPSI wafers used for PL  (Experimental Section). After He implantation to a fluence of $5   \times 10 ^{13} \, \mathrm{cm^{-2}}$, the CL spectrum has  the  same peak position and spectral width as the PL spectrum  of Fig.~\ref{fig1}b, demonstrating that electrons can indeed efficiently excite the $\mathrm{V_{Si}}$ defects. We then use HIM to write a barcode depicted in the insert of  Fig.~\ref{fig4}e. This barcode represents number 916,  corresponding to the spectral position  of one of the  $\mathrm{V_{Si}}$ zero-phonon lines in 4H-SiC \cite{10.1038/ncomms8578}. The barcode has been created using single pixel lines with a fluence of $70 \, \mathrm{ions/nm}$ at an energy of $25  \, \mathrm{keV}$ using a ion cuirrent of $0.5 \, \mathrm{pA}$.

The barcode retrieved with CL is presented in Fig.~\ref{fig4}c. To analyze the spatial resolution, we integrate the signal along the bars and plot the line scan  across the bars,  as shown in Fig.~\ref{fig4}e. From the multi-gaussian fit, we find the full width at half maximum (FWHM) of the narrower bars to be about $420 \, \mathrm{nm}$, which represents the spatial resolution. This value is in a good agreement with expected lateral spatial resolution given by the combined He and electron straggling $( \sigma_{\mathrm{HIM}}^2 + \sigma_{\mathrm{CL}}^2 )^{1/2} \approx 360 \, \mathrm{nm}$. 

According to Fig.~\ref{fig4}a and b, the spatial  resolution can be significantly improved if the creation depth of the $\mathrm{V_{Si}}$ defects is limited to $50 \, \mathrm{nm}$. This can be realized using He ions with lower energy in combination with thinner SiC epitaxial layers. The lateral straggling of He ions and electrons is limited in this case to $30 \, \mathrm{nm}$ and  $40 \, \mathrm{nm}$, respectively, resulting in a total spatial resolution of $50 \, \mathrm{nm}$. It corresponds to the storage density for  a single layer approaching  $300 \, \mathrm{Gbit / in^2}$, which is comparable to the record areal density of magnetic tapes  \cite{10.1109/tmag.2021.3076868} but with much longer storage time and without need of data migration. 

We note that the maximum density of optically active $\mathrm{V_{Si}}$ defects is $10 ^{16} \, \mathrm{cm^{-3}}$ \cite{10.1103/PhysRevApplied.13.044054}, while for higher concentration the presence of non-radiative recombination channels  due to irradiation-induced crystal damage suppresses the emission. For such a volume density, the average distance between single $\mathrm{V_{Si}}$ defects is $46 \, \mathrm{nm}$. Therefore in order to achieve the aforementioned areal storage density of $300 \, \mathrm{Gbit / in^2}$, the single bit should be stored in single $\mathrm{V_{Si}}$ defect. The creation yield of single $\mathrm{V_{Si}}$ defects using He ions with an energy of $6 \, \mathrm{keV}$  is $8.5\%$ \cite{10.1038/s41563-021-01148-3},  which corresponds to on average 12 He ions needed to create one $\mathrm{V_{Si}}$ center. Based on these numbers, the minimum energy required to store a single bit is estimated to be $\mathrm{10 \, fJ}$. This value does not take into account the energy consumption needed to keep the HIM running.  

The achievable writing speed is an important criterion for long term data storage. It is limited by the required fluence for the creation of the defects and the ion beam current. 
However, the latter is linked to the achievable spot size and can therefore not be increased indefinitely. 
If we assume a necessary fluence of $\Phi_{\mathrm{He}} =\SI{1.2e12}{\per\centi\meter\squared}$ and a bit area $\SI{1000}{\nano\meter\squared}$ according to the above presented findings we require the implantation of $\approx12$\,He ions. 
In case of HIM, we can use high currents in the range of $\SI{2}{\pico\ampere}$ as the required resolution can easily be reached. Under these conditions writing of a single pixel requires only $\SI{100}{\nano\second}$.
This results in a writing speed of \SI{10}{\mega\bit/\second}  (ignoring the required time for beam steering and refresh). This writing speed is faster than for other optical data storage media \cite{10.1364/optica.433765}. 

Optically active defects can also be achieved with other ions. A liquid metal ion source will allow to increase the current to a few \SI{10}{\nano\ampere}.
Provided that the spot size of $\SI{40}{\nano\meter}$ could be maintained, this nevertheless results in extremely short dwell times which will require new high speed beam blankers. 
Assuming such a fast blanker capable of blanking and unblanking the beam in less than \SI{10}{\nano\second} \cite{Klingner2015} a current of \SI{200}{\pico\ampere} is required to reach the required fluence $\Phi_{\mathrm{Ga}}$. In such a hypothetical FIB system writing speeds up to \SI{100}{\mega\bit/\second}  could be achieved.

The reading speed is directly limited by the emission rate of the $\mathrm{V_{Si}}$ defects. For the 4D  optical data storage described in Fig.~\ref{fig3}, the upper limit of the reading speed is about \SI{100}{\kilo\bit/\second}. It is much lower if single bits are stored in single $\mathrm{V_{Si}}$ defects, being in the range of only \SI{1}{\kilo\bit/\second}. Some other color centers in SiC demonstrate very strong PL, which can provide the reading speed of single defects exceeding \SI{1}{\mega\bit/\second} \cite{10.1038/nmat3806}. Alternatively, the count rate and, correspondingly, the reading speed  can be significantly improved by collecting PL from the $m$-face of the SiC crystal \cite{10.1038/s41534-022-00534-2} or collecting CL from SiC nanocrystals \cite{10.1063/1.4904807} deposited on a mirroring conducting surface.

\section{Conclusion} 

We demonstrate ultralong high-density data storage based on $\mathrm{V_{Si}}$ defects, which are created by focused ion beams. To retrieve the stored information, we excite the $\mathrm{V_{Si}}$  defects with a 785-nm laser used in CD drives and detect their PL. We then apply 2-bit and 8-bit grayscale encoding schemes based on the PL intensity to store text and images, respectively. The images are stored in two layers, and our estimations show that up to ten layers can be read without cross-talking. In this case, the projected storage density is in the range of tens $\mathrm{Gbit / in^2}$, which is comparable with HD DVD and Blu-ray disks. The extrapolation of the thermally activated PL decay time points at an extremely long storage time.  To overcome the diffraction limit, we then demonstrate that  CL can be used instead of PL to read bits encoded in optically active $\mathrm{V_{Si}}$ defects. To examine the spatial resolution, which determines the areal storage density, we write a barcode in 4H-SiC using HIM and read it using CL. Though the CL spatial resolution of about $360 \, \mathrm{nm}$ is better than in case of PL, it is limited by the straggling of ions and electrons during the write and read processes, respectively. By reducing the thickness of the recording layer to $50 \, \mathrm{nm}$, the spatial resolution can be significantly improved, providing the theoretical limit of the storage density about  $300 \, \mathrm{Gbit / in^2}$.

Our approach is not limited to SiC and can be extended to other materials with optically active defects, including 2D materials  \cite{10.1002/adfm.202103140}.

\section{Experimental Section} 

The 4H-SiC samples have been purchased from STMicroelectronics (formerly Norstel).  For the PL experiments, we use a high-purity semi-insulating (HPSI) 4H-SiC wafer.  For the CL experiments, we use an epitaxial p-type (doping level  $\mathrm{1 \times 10^{17} \, cm^{-3}}$) 4H-SiC layer with a thickness of  $\mathrm{250 \, nm}$, grown on an n-type  4H-SiC substrate.  A conductive substrate is needed to discharge the sample when it is exposed to the electron beam in CL experiments.  

The PL measurements are carried out at room temperature under vacuum conditions using a home-built confocal microscope. 
A continuous-wave 785-nm fiber-coupled laser (Thorlabs, LP785-SF20) is sent through a 785 nm band-pass filter with a 10 nm transmission window (AHF Analysentechnik AG, ZET785/10) and focused on the sample using a  microscope objectivewith NA=0.81 optimized for the NIR spectral range. The laser power is controlled using an in-line variable fiber optical attenuator (Thorlabs, VOA780-FC) and set to approximately $6 \, \mathrm{mW}$. Using a closed-loop nanopositioning system (attocube, ANPx311) anchored to the sample holder, 2-dimensional confocal PL raster scans of the spatially selective HIM patterns are performed. For the in-depth PL scans, the refractive index of SiC $\mathrm{n_{SiC}}$ = 2.6 is considered. The NIR emission from $\mathrm{V_{Si}}$ is collected by the same objective while the residuals of the laser light are cleaned-up by a set of two 800 nm (Thorlabs, FELH0800) long-pass filter. The PL is either fed into a broadband fiber-coupled superconducting nanowire single photon detector (Single Quantum) for spectrally integrated PL measurements or coupled into a spectrometer equipped with a Si-CCD (Andor Technology, iVac 324 FI) for wavelength-selective PL measurements.

The proton beam experiments are performed at the TIARA irradiation facilities of QST Takasaki, Japan. The $3  \, \mathrm{MV}$  single-ended particle accelerator is utilized to generate a stable focused beam with a typical size of $1 \times 1  \, \mathrm{\mu m^{2}}$, which was estimated at the end-station by secondary electron images on a copper mesh. The end station equips the precision 2-axis stage (Canon Precision) and controlled the position of the SiC wafer. The beam current through the sample holder was monitored during irradiation to register fluctuations and deviations from the intended current and fluence, while precise beam currents are evaluated before and after the irradiation experiment by a Faraday cup connected with a picometer.

All focused He ion beam irradiation has been performed in a Carl Zeiss Orion NanoFAB \cite{10.1116/1.4863676, 10.1007/978-3-319-41990-9}. The machine has several addons including a fast blanker capanble of the above mentioned ultra short pulse length \cite{NVision2007b}, which is also used for TOF-SIMS \cite{Klingner2015}.
If not noted otherwise a helium pressure of 2e-6mbar, a $10 \,\mathrm{\mu m}$ aperture, spot control number 4 and 25 kV acceleration energy have been used. To reduce the unintended irradiation by energetic neutral He atoms all irradiation patterns have been placed at least  $50 \,\mathrm{\mu m}$ (typically $100 \,\mathrm{\mu m}$) of axis of the ion optical column.

The commercially available optical detection system (delmic SPARC) used for the CL measurements is attached to a scanning electron microscope (scienta omicron NanoSAM Lab). The sample is mounted on a piezo controlled micropositioning system. The electron beam is directed through a micrometer-sized aperture of the parabolic mirror and focused on the sample surface down to a spot size below $\mathrm{10 \, nm}$. A CL signal is emitted as the continuous electron beam of $\mathrm{5 \, keV}$  ($\mathrm{1 \, nA}$) excites  $\mathrm{V_{Si}}$ at the focal point of the parabolic mirror.
The CL response is collected by the same parabolic mirror and either fiber-coupled (Thorlabs, FG300AEA) to a Si-APD (Thorlabs, APD440A) for spectrally integrated CL measurements or coupled into a spectrometer equipped with a Si-CMOS camera (Andor Technology, Zyla 5.5 sCMOS) for wavelength-selective CL measurements. 2D CL intensity maps are obtained by scanning the electron-beam laterally across the HIM pattern. To avoid contributions from other luminescent defects in the visible spectral range, the CL signal is long-pass filtered at  $\mathrm{830 \, nm}$ (Semrock, LP02-830RE-25). All CL measurements were carried out at room temperature under vacuum conditions.

\section*{Acknowledgments}
V.D. and G.V.A. acknowledge financial support from the W\"urzburg-Dresden Cluster of Excellence on Complexity and Topology in Quantum Matter (EXC 2147, DFG project ID 390858490). T.O. acknowledges the support from MEXT Q-LEAP JPMXS0118067395. M.Ho. thanks Helmut Schultheiss for the assistance with Blender by the preparation of schemtics in Fig.~\ref{fig1}a. G.H. and L.B. are members of the COST Action FIT4NANO CA19140 http://www.fit4nano.eu. H.K. carried out work at the Jet Propulsion Laboratory, California Institute of Technology, under a contract with the Nastional Aeronautics and Space Administration (80NM0018D0004). Support from the Ion Beam Center (IBC) at HZDR and TIARA irradiation facilities of QST Takasaki are gratefully acknowledged for ion implantation. 


%

\end{document}